\documentclass[letterpaper]{article}

\usepackage[T1]{fontenc}

\usepackage{geometry}
\usepackage{setspace}

\usepackage{achemso}

\usepackage{graphicx}
\usepackage{float}
\newfloat{scheme}{htbp}{los}
\floatname{scheme}{Scheme}
\floatname{chart}{Chart}
\newfloat{graph}{htbp}{loh}

\usepackage{chemformula} 
\usepackage[version = 4]{mhchem} 

\usepackage{dcolumn}
\usepackage{bm}
\usepackage{multirow}
\usepackage{amsmath}
\usepackage[utf8]{inputenc}
\usepackage{physics}
\usepackage{tabularx}
\usepackage{color}
\usepackage{appendix}
\usepackage{enumitem}
\usepackage{comment}
\usepackage[caption=false]{subfig}
\usepackage{authblk}
\author[1]{Jeremy Canfield}
\author[2]{Dominika Zgid}
\author[1]{James K. Freericks*}
\affil[1]{Department of Physics, Georgetown University, 37th and O Sts. NW, Washington, DC 20057 USA}

\affil[2]{Department of Chemistry, University of Michigan}

\title{Low Depth Unitary Coupled Cluster Algorithm for Large Chemical Systems}
\date{*Email: james.freericks@georgetown.edu}

\begin{document}
\maketitle

\begin{abstract}
The unitary coupled cluster (UCC) algorithm is one of the most promising implementations of the variational quantum eigensolver for quantum computers. However, for large systems, the number of UCC factors leads to deep quantum circuits, which are prohibitive for execution on quantum hardware. To address this, circuit depth can be reduced at the cost of more measurements with a Taylor series expansion of UCC factors with small angles, while treating the large-angle factors exactly. We implement this approach to quadratic order (qUCC) for systems with strong correlations and systems where conventional methods like coupled cluster (CC) with low excitation levels fail, but UCC and qUCC perform well. We study hydrogen chains and the BeH$_2$ molecule that allow us to change the degree of strong correlation due to geometrical distortions. We show, via a dramatic increase in number of factors able to handle exactly, a systematic convergence of these results as more exact UCC factors are included in the calculations---the hardest to converge regime is in the crossover from weak to strong coupling. In all cases the total number of UCC factors needed to be treated exactly is much less than the total number of UCC factors available (typically about one-third to one-half of the total number of factors).

\end{abstract}
\maketitle

\section{Introduction}

The variational quantum eigensolver for quantum chemistry electronic structure calculations has been proposed as an important application for early quantum computers. Indeed, early implementations of the approach showed great promise~\cite{VQE}. However, as research progressed, it became clear that the depth of these circuits, even for relatively simple systems, grows much too quickly to be implemented in near term hardware~\cite{NISQ,qUCC_imp,trapped_ion_sim, dejong, ionq, UCC_imp,UCC_imp2}. In fact, the total circuit depth is challenging even for soon to be available fault-tolerant machines.

Conventional coupled cluster (CC) is characterized by the amplitudes of the excitation operators. In the unitary coupled cluster (UCC)~\cite{UCC_1,UCC_2,UCC_3} approach to the variational quantum eigensolver, one typically employs a factorized form for the UCC operator, where each excitation appears separately as a factor with an associated angle. The UCC angle is related to the corresponding CC amplitude, to first order, by a $\theta\to\tan(\theta)$ relationship~\cite{JKF_CC}. It is common to order the UCC doubles factors according to their importance in the second order M\o ller Plesset (MP2)~\cite{MP2} perturbation theory. When such calculations are performed, one finds that only a small number of angles are large and most are small. They are not so small that they can be neglected, but they are not large enough to require exact treatment. This motivates a hierarchical approach---we treat the large angle terms exactly with UCC factors included on the quantum hardware and, for the remaining factors, we perform a quadratic Taylor series expansion of the energy about the origin for the small angles and about the optimized values for the large angles. This allows us to implement the calculations with much shallower depth circuits, and use row reduction, rather than nonlinear optimization to determine the best small angles' values. This approach naturally extends to include higher-order terms beyond singles and doubles as well, if those terms also have small angles. This allows for refinement of calculations performed with lower-rank UCC factors.

This expansion for the small-angle UCC factors is called quadratic UCC or qUCC. It has already shown promise~\cite{qUCC_1}. The earlier work was severely limited in how many UCC factors could be handled exactly on the classical computer---only about 30 factors. In this work, we have made the algorithm much more efficient, so that we can include about 300 exact factors, allowing us to carefully check the convergence of the approach with respect to the number of exact angles included. We find that it converges well to the full UCC result with only a small fraction of all singles and doubles terms being treated exactly. This significantly reduces computational time and trades reduced circuit depth for more measurements, a preferred tradeoff for near-term machines (including fault-tolerant ones).  In essence, the quantum computer is primarily needed for its memory to store the quantum state generated by the ansatz and the nonlinear optimization procedure is restricted only to the exact angles, while the Taylor expanded angles are determined using linear algebraic methods.

In this work, we focus our calculations on classical machines with systems that can be solved via full configuration interaction (FCI) calculations, so that we have exact results for benchmarking. We examine a number of challenging cases where traditional CC approaches fail. The main result we find is that there is a systematic way to determine how many exact UCC factors are needed to achieve convergence, and that once converged, the approach is quite accurate in determining the optimized energy and ground state with a significantly more efficient algorithm than working with a full UCC calculation.

\section{Theory and Method}
The UCC method employs an exponential quantum state ansatz of the form 
\begin{equation}\label{eq::cc_equation}
    \ket{\Psi_{\text{UCC}}} = \exp(\hat{\sigma})\ket{\Psi_0},
\end{equation}
 where $\ket{\Psi_0}$ is some initial single reference state (often the Hartree-Fock ground state), and the operator $\hat{\sigma}$ is 
 \begin{equation}
     \hat{\sigma} = \hat{T}-\hat{T}^\dagger,
 \end{equation}
with $\hat{T}$ being a sum over possible excitation operators (singles, doubles, etc.):

\begin{equation}
    \hat{T} = \underbrace{\sum_i^{occ}\sum_a^{vir}\theta_i^a \hat{a}^\dagger_a \hat{a}_i}_{\text{singles}} +\underbrace{\sum_{i>j}^{occ}\sum_{a>b}^{vir}\theta_{ij}^{ab} \hat{a}^\dagger_a\hat{a}^\dagger_b \hat{a}_i \hat{a}_j }_{\text{doubles}}+ \underbrace{\cdots}_{\text{higher}}
\end{equation}
where $a,b...$ represent virtual orbitals and $i,j...$ represent occupied orbitals in the reference state. To make this notation simpler, we represent an $n$-fold excitation operator as $\hat{a}_{ij...}^{ab...}=\hat{a}^\dagger_a\hat{a}^\dagger_b ... \hat{a}_i \hat{a}_j...$. The symbol $\theta_{ij...}^{ab...}$ is the angle of the $n$-fold excitation that measures its strength with respect to the chosen reference. 
In the UCC formalism, the expression from Eq.~\ref{eq::cc_equation} can be approximated by its factorized form, given by 
\begin{equation}
    \ket{\Psi_\text{UCC}} \approx \prod_k e^{\theta_k \sigma_k}\ket{\Psi_0} = \prod_{i>j...}^{occ}\prod_{a>b...}^{vir}\exp[\theta_{ij...}^{ab...}(\hat{a}_{ij...}^{ab...}-\hat{a}_{ab...}^{ij...})]\ket{\Psi_0}= \prod_{i>j...}^{occ}\prod_{a>b...}^{vir}U_{ij...}^{ab...}\ket{\Psi_0} ,
\end{equation}
which we use in this work. Note that the factorized form is different from the unfactorized form, but because we use the variational principle to optimize the exact parameters, it usually has enough expressiveness that one can find optimized states within this more restrictive form of the ansatz. 

There is a hidden SU(2) symmetry in this factorized form~\cite{UCCSU2} which allows us to write a generalized UCC excitation/de-excitation operator, $U$, in the simpler form
\begin{align} 
\label{eq:udef}
        U_{ij...}^{ab...} = \exp[\theta_{ij...}^{ab...}(\hat{a}_{ij...}^{ab...}-\hat{a}_{ab...}^{ij...})] 
    =1 + \sin(\theta)(\hat{a}_{ij...}^{ab...}-\hat{a}_{ab...}^{ij...})+
    \\ \nonumber
    (\cos(\theta)-1)[\hat{n}_a\hat{n}_b...(1-\hat{n}_i)(1-\hat{n}_j)...+(1-\hat{n}_a)(1-\hat{n}_b)...\hat{n}_i\hat{n}_j],
\end{align}
where $\hat{n} = \hat{a}^\dagger\hat{a}$ is the number operator. This is essentially an operator generalization of the Euler formula. When each operator is applied, there are two options: either the operator is unable to cause an excitation or de-excitation and acts as the identity, or it can excite/de-excite some electron(s), resulting in a superposition of the quantum state where the previous state is multiplied by $\cos\theta$ and an additional state weighted by $\sin\theta$ is added to it. This allows for a simple strategy using a binary tree structure for generating the quantum state given some collection of UCC factors. Note that UCC factors $U_{ij...}^{ab...}$ are not equivalent to traditional CC amplitudes~\cite{JKF_CC}. But similar to traditional CC, where an increased number of amplitudes results in improved accuracy, in the factorized UCC, an increased number of factors (with their associated angles instead of amplitudes) leads to an improved agreement with the exact result.  


Equation~(\ref{eq:udef}) also allows us to derive a simple formula for the derivative of a factor $U$ with respect to the variational parameter, the angle $\theta$:
\begin{equation}
    \frac{dU}{d\theta} = \cos(\theta)(\hat{a}_{ij...}^{ab...}-\hat{a}_{ab...}^{ij...})-\sin(\theta)[\hat{n}_a\hat{n}_b...(1-\hat{n}_i)(1-\hat{n}_j)...+(1-\hat{n}_a)(1-\hat{n}_b)...\hat{n}_i\hat{n}_j].
\end{equation}

On a quantum computer, there are a number of different ways to implement each UCC factor. For lower-rank terms, often a direct implementation using parity-based circuits works best, while for higher-ranks, approaches based off of the generalized Euler formula will work better~\cite{LuogenUCC,evangelista}. No matter how it is implemented, the circuit for a large number of UCC factors becomes very deep, cannot be implemented on near term hardware, and may be problematic even for fault-tolerant hardware. Thus, to limit circuit depth, we implement an approximate scheme based on a Taylor expansion in the energy, called the qUCC method. Here, the first approximation to the UCC energy is found from the expectation value of the Hamiltonian 
\begin{equation}
    E_\text{UCC} = \bra{\Psi_\text{UCC}}\hat{H}\ket{\Psi_\text{UCC}} = \bra{\Psi_{0}}\hat{U}_1^\dagger...\hat{U}_L^\dagger|\hat{H}|\hat{U}_L...\hat{U}_1\ket{\Psi_{0}}
\end{equation}
for the $L$ UCC factors that are treated exactly. In the limit where $L\to N$, the total number of factors, this is the conventional factorized UCC. In principle, much of the initial optimization can be done on a classical computer (say for a few hundred or more of the most important UCC factors), leaving fewer optimization steps to be carried out on the noisy quantum hardware. Then, to quadratic order, we  improve the expectation value of the energy by expanding it according to  
\begin{equation}\label{eq:en_taylor}
    E \approx \bra{\Psi_{0}}\hat{H}\ket{\Psi_{0}} + \sum_k b_k(\theta_k'-\theta_k) + \frac{1}{2}\sum_{k,m} A_{km}(\theta_k'-\theta_k)(\theta_m'-\theta_m),
\end{equation}
where 
\begin{eqnarray}
    b_k = \frac{\partial\expval{H}}{\partial\theta_k} = 2\text{Re}\bra{\Psi_{0}}\hat{U}_1^\dagger...\hat{U}_N^\dagger|\hat{H}|\hat{U}_N...\frac{\partial U_k(\theta_k)}{\partial\theta_k}...\hat{U}_1\ket{\Psi_{0}},\label{eq:bdef} \\
    A_{km} = \frac{\partial^2\expval{H}}{\partial\theta_k\partial\theta_m}  = 2\text{Re}\bra{\Psi_{0}}\hat{U}_1^\dagger...\frac{\partial \hat{U}^\dagger_m(\theta_m)}{\partial\theta_m}...\hat{U}_N^\dagger|\hat{H}|\hat{U}_N...\frac{\partial \hat{U}_k(\theta_k)}{\partial\theta_k}...\hat{U}_1\ket{\Psi_{0}} \\ \nonumber + 2\text{Re}\bra{\Psi_{0}}\hat{U}_1^\dagger...\hat{U}_N^\dagger|\hat{H}|\hat{U}_N...\frac{\partial \hat{U}_k(\theta_k)}{\partial\theta_k}...\frac{\partial \hat{U}_m(\theta_m)}{\partial\theta_m}...\hat{U}_1\ket{\Psi_{0}},\label{eq:adef}
\end{eqnarray}
and $\theta_k',\theta_m'$ are the exact angles around which we perform the expansion. Note that $\overrightarrow{b}$ has $N$ elements, and $\bf{A}$ is an $N\cross N$ matrix.
This expansion allows us to include additional UCC factors that will have small, but nonnegligible angles.
This expansion is accurate when the Taylor-expanded amplitudes remain small. Hence, we use an initial second order M\o ller-Plesset (MP2) calculation to group the angles into small and large subsets. Then, 
we apply the factors associated with large angles exactly to construct an initial UCC state. We perform an optimization over this  subset of variational parameters to obtain
\begin{equation}\label{eq:large_factors_state}
    \ket{\Psi_\text{UCC}^0} = \prod_l^L 
    \hat{U}_l(\theta_l)\ket{\Psi_0},
\end{equation}
where $l$ indexes the large angles and $\theta_l$ are the parameters explicitly optimized (in this work we use the BFGS optimization routine in SciPy)~\cite{BFGS}. This reduces circuit depth by only running the nonlinear optimizer over the small subset of parameters that are large, and gives us a good starting point to refine with the Taylor series expansion. The key is that the Taylor series expansion is not about the origin in parameter space but about this optimized state that contains large-angle factors. Note that if $L$ is small enough, the initial optimization can be done on a classical computer. If $L$ is too large, then an initial optimization will still be done classically to properly seed the initial values for a quantum calculation to fully optimize all $L$ large angles. 

Now, we can plug this partially optimized reference state into Eq.~(\ref{eq:en_taylor}) to obtain a quadratic approximation of the UCC energy relative to this reference state. Note that the vector $\vec{\theta}$ used in calculating the $b$ vector and $A$ matrix is of the form
\begin{equation}
\vec{\theta}=\big(\underbrace{\theta_1,\theta_2,\cdots\theta_L}_{L~\text{terms}},\underbrace{0,0,\cdots,0}_{N-l~\text{terms}}\big).
\end{equation}
Note that the initial values of the first $L$ angles are found from the initial nonlinear optimization step. In Eqs.~(\ref{eq:bdef}) and (\ref{eq:adef}), the $\hat{U}_i$ factors depend on $\theta_i$ for the first $L$ factors, but are $\hat{I}$ for the remaining $N-L$ factors. 
We then perform a linearized optimization step by taking the derivative of Eq.~(\ref{eq:en_taylor}), and setting it equal to zero:
\begin{equation}
    \frac{\partial E}{\partial \theta_i} = b_i + \sum_j A_{ij}\theta_j = 0.
\end{equation}
 This forms a system of simultaneous linear equations,
\begin{equation}\label{eq:system}
\bf{A}\cdot\overrightarrow{\theta} = -\overrightarrow{b}.
\end{equation}
Solving this system yields the optimal angles, both large and small. This is a key point: we optimize all the angles, but only the first $L$ angles with a nonlinear optimizer (the initial optimization). The final optimization on the full parameter space is done by a matrix inversion, a much less costly operation. We then compute the final result for the qUCC algorithm by evaluating Eq.~(\ref{eq:en_taylor}) with the new optimized values for both large and small angles. 

\begin{figure}
    \centering
    \includegraphics[width=0.95\columnwidth]{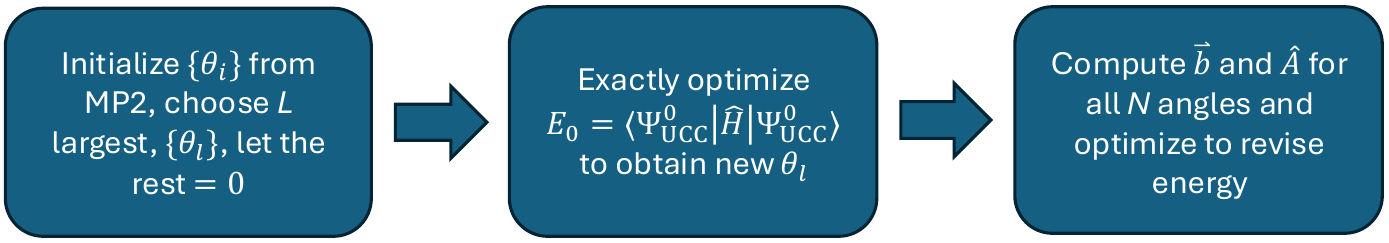}
    \caption{Flowchart outlining the procedure of the qUCC algorithm.}
    \label{fig:code_flow}
\end{figure}

We only consider singles and doubles excitation factors, and the initial values of their angles are estimated by running an MP2 calculation. Since all singles angles are zero in MP2~\cite{MP2}, so we initially choose only doubles angles to be large. However, after solving the system in Eq.~(\ref{eq:system}), we then re-evaluate if any new angles, including singles, are large, and we promote them to the group of large angles, repeating the algorithm for this new set of large angles. For the general case, one can include higher rank factors as well both in the large angle and small angle categories. We did not need to do this here.  In this work, we instead study the behavior of the qUCC algorithm as we add more factors, so we select the largest $L$ factors initially using the MP2 scheme. Since, in the systems we consider here, there are no more than a few dozen singles, we have a fairly liberal criterion for promoting them to the large group: if their magnitude rises above $10^{-4}$ for any geometry we study. 

One potential concern for this process occurs when the matrix $\bf{A}$ has zero, or near zero determinant. Then the inverse of $\bf{A}$ is not well-defined, and the row-reduction solver may fail to produce accurate solutions. We did not encounter this issue in the systems we analyze here, but previous work~\cite{qUCC_1} has addressed how to deal with this if it does arise---one simply works with a singular value decomposition to solve the row-reduction problem.

The qUCC algorithm provides a practical method for electronic structure calculations on a quantum computer. In the presence of noisy data, limiting the number of variables to optimize, and as a result, the circuit depth, is a major benefit, and as we will see, this method allows for a dramatic reduction in number of factors we need to treat exactly, when compared to a full UCC approach. The tradeoff, then, is the need to take many more measurements in calculating the $b_k$ vector and the $A_{km}$ matrix. However, these measurements only involve calculating derivatives (and second derivatives) with respect to variational parameters once, whereas in the standard optimization of a traditional UCC calculation, one would  compute derivatives with respect to the variational parameters at each optimization step. Often, many optimization steps are required. This implies that qUCC will substantially reduce both the circuit depth and the overall compute time, so long as the number of exact factors is small. 

Finally, we remark briefly on the conventional CC~\cite{CC_1,CC_2} method to highlight a few differences that will become apparent in our analysis. Conventional CC is a non-unitary exponential ansatz of the form
\begin{equation}
    \ket{\Psi_\text{CC}} = \exp(\hat{T})\ket{\Psi_0}.
\end{equation}
The CC energy is obtained by evaluating
\begin{equation}\label{eq:ECC}
    \bra{\Psi_\text{CC}}e^{-\hat{T}}\hat{H}e^{\hat{T}} \ket{\Psi_\text{CC}},
\end{equation}
with the $\hat{T}$ being the same as above for the UCC, but now without the accompanying Hermitian conjugate. It optimizes the amplitudes by zeroing out the Hartree-Fock column of the transformed Hamiltonian matrix. The benefit of the method is that the nested commutators between the Hamiltonian and the $\hat{T}$ operator terminate at fourth order, so one can perform the calculation without ever constructing the quantum state. This termination does not occur for UCC, which is why the full state must be prepared, and is why a quantum computer is needed, because conventional computers will run out of memory too soon. However, since $-\hat{T}$ is not the Hermitian conjugate of $\hat{T}$, the expression in Eq.~(\ref{eq:ECC}) does not produce a variational estimate of the energy.  Hence, it is possible to obtain results that go below the exact result of FCI. Notably, CC, due to its nonvariational nature, is known to fail and can yield energies that lie below FCI in regimes with strong correlations. On the other hand, UCC methods, which are variational, have shown promise in providing accurate results in these regimes.~\cite{strong_corr} We apply the qUCC algorithm to some simple examples that display strong correlation where UCC methods avoid common CC failures.

\section{Algorithm Methodology}
We comment on some of the details of our algorithm. As we noted before, previous work~\cite{qUCC_1} on this method hit a hard limit for number of large factors possible to include at about 30. As we show in this work, we were able to push this an order of magnitude further, which allows us to establish that the qUCC method converges with just a fraction of all singles and doubles factors. The major improvement in our algorithm to drive this efficiency was the implementation of a procedure for constructing the action of the electronic Hamiltonian on the UCC state that is in line with standard quantum chemistry practice. In the previous work, out of convenience, the entire Hamiltonian matrix, or a matrix in a subspace of all relevant determinants, was constructed.  This matrix grows rapidly as more exact factors are included, and constructing it is inefficient when many elements are zero anyways. Our update is to just compute the many-body Hamiltonian matrix elements on the fly from the one and two-particle integrals brings the implementation of the qUCC algorithm in line with traditional efficient quantum chemistry approaches.

The electronic Hamiltonian in second quantization is 
\begin{equation}
    \hat{H} = \sum_{pq}h_{pq}\hat{a}^\dagger_p \hat{a}_q +\frac{1}{2}\sum_{pqrs}h_{pqrs}\hat{a}^\dagger_p\hat{a}^\dagger_q\hat{a}_r\hat{a}_s,
\end{equation}
where $h_{pq}$ and $h_{pqrs}$ are the one- and two-electron integrals. After finding the energy eigenvalues of the electronic Hamiltonian, we add in the ionic energy corresponding to the repulsion of the ion cores, to determine the total energy.
For UCC and qUCC, we need to calculate the action of this Hamiltonian on various vectors, namely, the quantum state and first and second derivatives of that quantum state with respect to variational parameters. Constructing and storing the full Hamiltonian quickly becomes prohibitive, and even constructing partial Hamiltonian matrices is prohibitive in systems beyond very small sizes. In this work, use traditional quantum chemistry methods in the following way.

Whenever an action of the Hamiltonian on some vector $\ket{\psi}$ is needed, the result will be a vector of the same size of $\ket{\psi}$, that is 
\begin{equation}
    \hat{H}\ket{\psi} = \ket{\psi_H}.
\end{equation}  
Each element of the vector $\ket{\psi_H}$ is a sum of the contributions of various elements $h_{pq}$ and $h_{pqrs}$ multiplying the terms that the quadratic or quartic operators map the product state components of $\ket{\psi}$ to after the action of the $\hat{H}$ operator. By efficiently implementing the Slater-Condon rules, each element of the resultant action vector is determined by calculating the minimum number of Hamiltonian elements on the fly, thus requiring no large memory resources to store the huge many-body Hamiltonian matrix, nor the cost of constantly accessing that memory, all while eliminating unnecessary calculations. Instead, we store and access $h_{pq}$ and $h_{pqrs}$, which are much smaller than the full Hamiltonian (100 and 1540 elements respectively for H$_{10}$, compared to the $\times 10^9$ total elements of the Hamiltonian), and due to symmetry of the one- and two-electron integrals, their number grows much more slowly than the many-body Hamiltonian matrix as the system size grows. 
For our purposes, this implementation was made easy by the ``contract\_2e'' routine in the PYSCF software library\cite{pyscf}, which efficiently gives the action of a Hamiltonian on a vector, given the inputs of the initial vector and the one- and two-electron integrals. 

This improvement to the algorithm was particularly important in allowing us to implement more large factors for the initial optimization. If one has $L$ large factors, as one can see from Eq.~(\ref{eq:large_factors_state}), the quantum state can have as many as $2^L$ determinants with non-zero coefficients in it, as each factor has the possibility to bifurcate the quantum state. In reality, the amount of determinants present grows more slowly, as many determinants are duplicated in the sequential application of excitation and de-excitation factors, and many factors cannot operate on many determinants. Still, the more factors there are, the larger the determinant subspace of the calculation, and the more Hamiltonian elements one would need to calculate and store with a more naive approach. This number grows further when one considers too that derivatives and second derivatives of the quantum state with respect to the angles of the factors will have additional determinants with non-zero coefficients. Thus, by circumventing the need to construct and store a full many-body Hamiltonian matrix, or even a partial Hamiltonian matrix, and instead calculating only the necessary many-body matrix elements on the fly, we have removed a significant computational limit in treating additional factors as large, resulting in our ability to study the behavior of the qUCCSD method for many more exact factors than previously.

Ultimately, as we show and discuss below, the computational limit we encounter in this work is a runtime, not memory, limit. In principle, the qUCC algorithm is not FCI limited. It is limited by the growth of the quantum state needed for the $L$ exactly treated factors, the number of two-body integrals $h_{pqrs}$, and the number of virtual factors included, as represented by the size of the $\bf{A}$ matrix. In practice, for a large system, many determinants will be zero in the quantum state due to the limited number of factors applied exactly, so the size of the quantum state will grow slower than the FCI dimension, and this method should be extendable well beyond the limits of FCI, or of the memory limitations for larger systems given by their Hilbert-space dimension (if the $L$-factor UCC state needed has a size significantly smaller than the full Hilbert-space dimension). 

\section{Results and Analysis}
Throughout this work, one- and two-electron integrals, as well as results for methods such as MP2 and CCSD, and other relevant chemical values, are obtained from the PYSCF software~\cite{pyscf}. All the UCC values were obtained from in house codes developed by the authors of this paper.
\subsection{Linear Hydrogen Chains}
We begin with linear hydrogen chains: H$_{6}$, H$_8$, and H$_{10}$ in the minimum STO-3G basis. By stretching the bond length between the atoms, the molecule enters a strongly correlated regime. This regime is of particular interest because many weakly correlated methods, such as traditional CC with low excitations, are known to fail here. We examine the energy associated with bond lengths from $0.4$~\AA~ to $2.4$~\AA, with strong correlations expected when we get sufficiently far from the equilibrium bond length. Studying these three simple systems together allows us to analyze the scaling of our method with respect to the number of exact factors that are included in the expansion. From the Hartree-Fock reference state, H$_6$ in the minimal basis has 18 singles and 99 doubles for a total of 117 UCCSD factors. H$_8$ has 32 singles and 328 doubles (360 total), and H$_{10}$ has 50 singles and 825 doubles (875 total). 

\begin{figure}
    \centering
    \subfloat{\includegraphics[width=0.45\columnwidth]{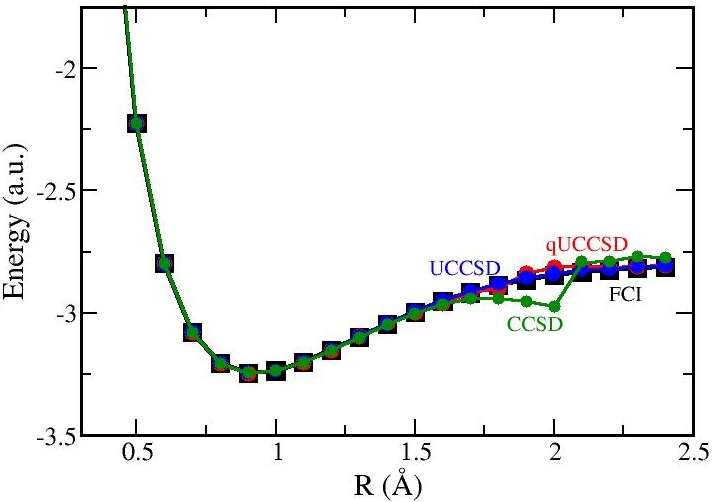}}
    \qquad
    \subfloat{{\includegraphics[width=0.45\columnwidth]{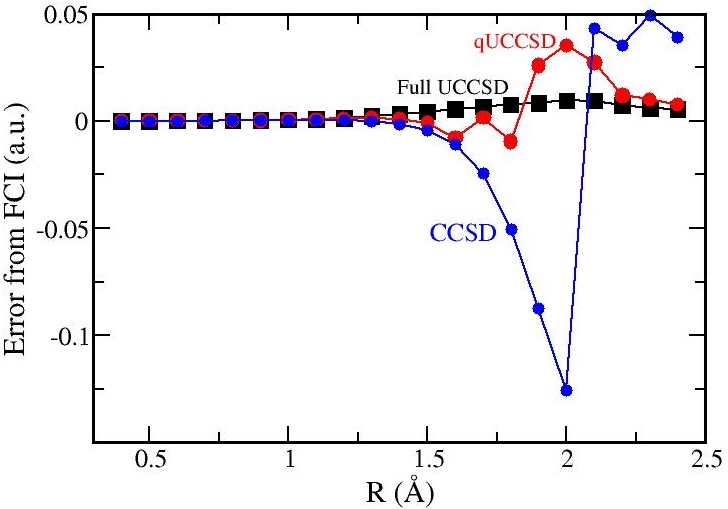} }}

    \caption{{\em Left}: total energy for the H$_6$ linear chain as a function of bond length. {\em Right:} the error from FCI, as $E-E_\text{FCI}$. Here, 30 exact factors are used for qUCCSD. We observe that qUCCSD performs similarly to both UCCSD and CCSD for the equilibrium bond length while outperforming CCSD in the presence of strong correlations. The poorest agreement is observed in the crossover region around 2 \AA. Note that while UCC is always variational, qUCC need not be due to the truncation of the quadratic expansion. For smaller $L$ cases, such as shown here, it does go non variational in the intermediate coupling regime.}
    \label{fig:Hchain_1}
\end{figure}

\begin{figure}
    \centering
    \subfloat{\includegraphics[width=0.45\columnwidth]{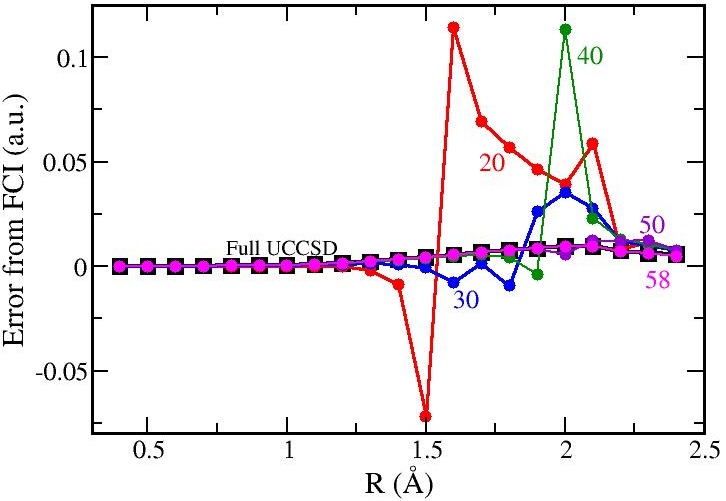}}
    \qquad
    \subfloat{{\includegraphics[width=0.45\columnwidth]{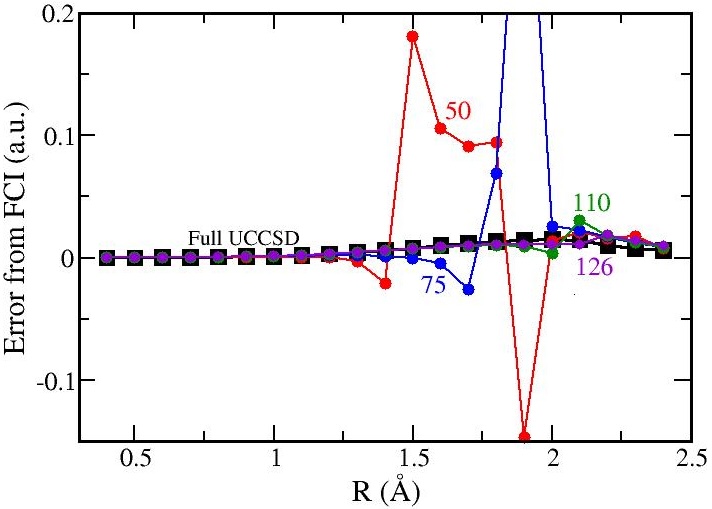} }}
    \\ \subfloat{{\includegraphics[width=0.45\columnwidth]{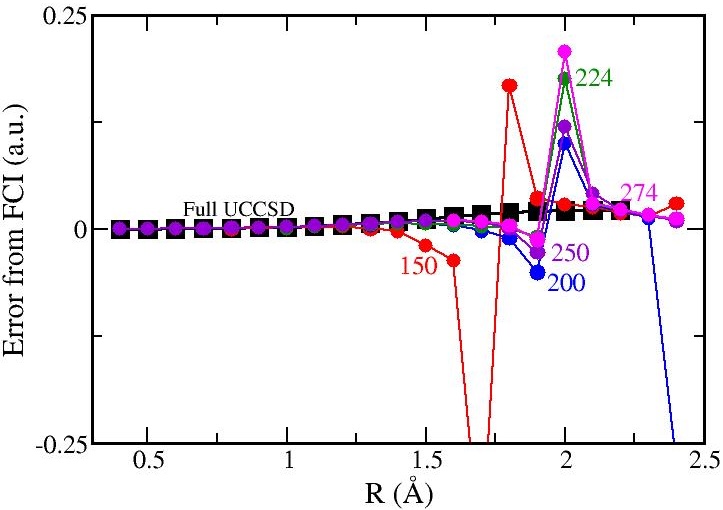} }}

    \caption{Energy difference from FCI for the 6- (left), 8- (right), and 10- (bottom) atom hydrogen chains. We illustrate how the accuracy improves with more exact factors, eventually converging to the full UCC result. Convergence is shown for H$_6$ and H$_8$, where we use a small fraction of the total doubles factors, and  a small number of singles (58 and 126 factors, respectively).  We cannot achieve full convergence 
    For H$_{10}$, with the largest number of factors listed (which is about one-fourth of the total number available), we do not achieve a perfect convergence to the UCC result. We cannot increase the number of exact factors due to the computational constrains, however, for this case, even in the intermediate regime, one observes fast convergence to the UCC result as the number of the exact factors is increasing.  }
    \label{fig:Hchain_2}
\end{figure}

The results for these linear chains are summarized in Figures~\ref{fig:Hchain_1} and \ref{fig:Hchain_2}. In Fig.~\ref{fig:Hchain_1}, we show the energy curve and error from FCI for qUCCSD (using 30 exact factors) and several other methods for the H$_6$ molecule. We see that even for a small number of exact factors, qUCCSD is accurate for weak correlations. As the correlations increase, qUCCSD and CCSD both fail, but this allows us to analyze the effect of adding more exact factors. Note that for H$_8$ and H$_{10}$, the total energy curves are similar in shape to H$_6$, so we focus on the error from FCI for the remainder of our analysis, and the energy curves for H$_8$ and H$_{10}$ are shown in the SI.

In Fig.~\ref{fig:Hchain_2}, we show the effect of adding more exact factors for different stretches. We see that only a small number of  factors are needed to achieve accurate results in the weak correlation regime. At the largest stretches, we find that a relatively small number of exact factors are needed, as well. It is the intermediate regime, where strong correlations first begin to enter while the weak correlations are still present, that proves to be the most difficult region to achieve accurate results. This implies that more factors are needed in this intermediate regime. Notably, in all cases we treat, this regime lines up with the regime in which CC fails, but unlike CC, we have 

For H$_6$ and H$_8$, we find that adding more exact doubles gets  close to the full UCCSD results, and including some singles as exact factors converges to results almost identical to full UCCSD. This is with a steep reduction of the total factors needed: for H$_6$, 58 (50 doubles and 8 singles) vs.~117 total factors and for H$_8$, 126 (110 doubles and 16 singles) vs 360 total. We reached our computational limit in the H$_{10}$ case, we could converge the full UCCSD results only up to $2.2$\AA, and qUCCSD only fully converged in the weak coupling regime. This computational limit was runtime in nature, and since we had convergence for H$_{10}$ in all but a small region, we are not checkpointing the code and attempting to run longer than system wall clock limits. The demonstration of convergence by increasing the number of exact factors is clear as is the fact that we always need significantly fewer factors to achieve this convergence, even for computationally hard cases. Of course, reaching this computational limit on classical hardware that is memory-limited is inevitable, which is why we need to use quantum hardware to examine  larger systems. Taking these three results all together, not only does qUCCSD offer a notable reduction in circuit depth versus the full UCCSD, but the method scales slower than system size does, with convergence at about half the total factors for H$_6$ and just over a third of total factors for H$_8$. Although we do not achieve convergence in the intermediate regime for H$_{10}$, these problematic points are limited to a small range of bond lengths even at less than a third of the total parameter space treated exactly.  

\subsection{Beryllium Insertion}
The next case we consider is the insertion of a beryllium atom between two fixed hydrogen atoms. We again use the minimal STO-3G basis, resulting in a system with 204 UCCSD factors: 180 doubles and 24 singles. We fix the hydrogen atoms along the $x$-axis at $\pm1$\AA, the beryllium is located at the origin along the $x$-axis, and we move the beryllium atom along the $y$-axis from $-2.4$\AA~to $2.4$\AA, which, as one would expect, is symmetric about the $x$-axis. We note that if one looks at the FCI for the ground state and first excited state for this problem, there is a level crossing near $\pm 2$\AA ~and another near $\pm 1.5$\AA. Between these crossings, the ground state is described by a multi-reference state, as opposed to the single reference of the Hartree Fock ground state. 
However, the first excited state is described by the Hartree Fock reference state throughout this region.
So, for this demonstration, we track the state described by the single reference HF state, which is the ground state everywhere except for the regions between the crossings, where it is the first excited state. In other words, we adiabatically track the single reference state from the HF reference throughout the entire region, regardless of whether it is the ground state or not.

\begin{figure}
    \centering
    \subfloat{\includegraphics[width=0.45\columnwidth]{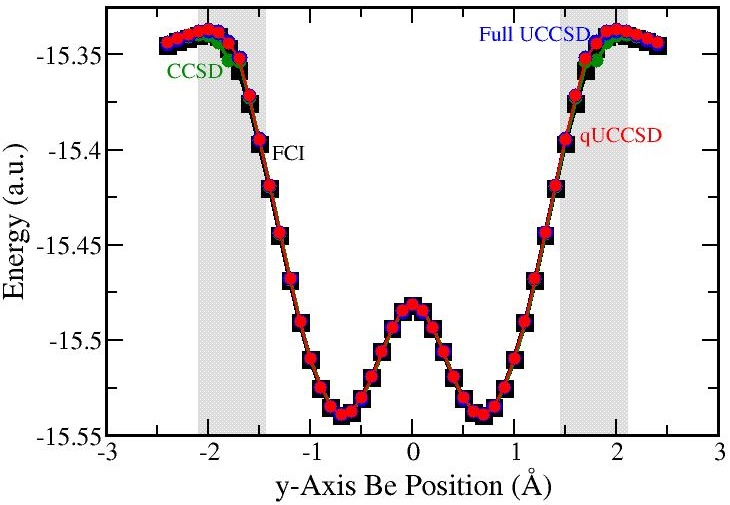}}
    \qquad
    \subfloat{{\includegraphics[width=0.45\columnwidth]{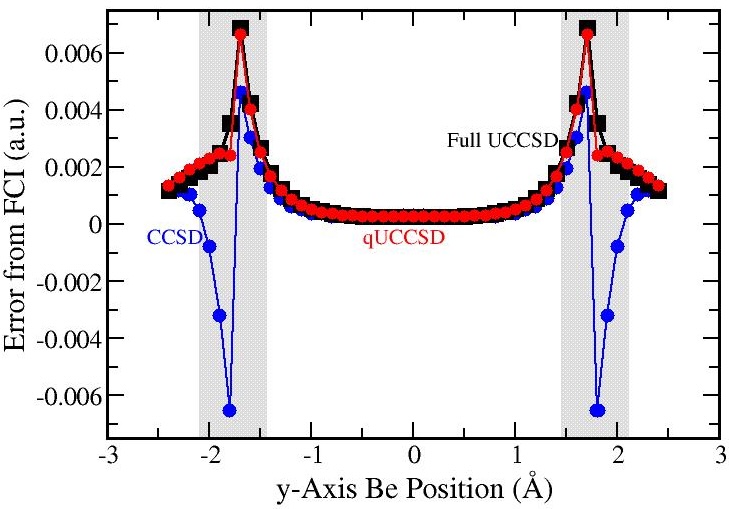} }}
    \\ \subfloat{{\includegraphics[width=0.45\columnwidth]{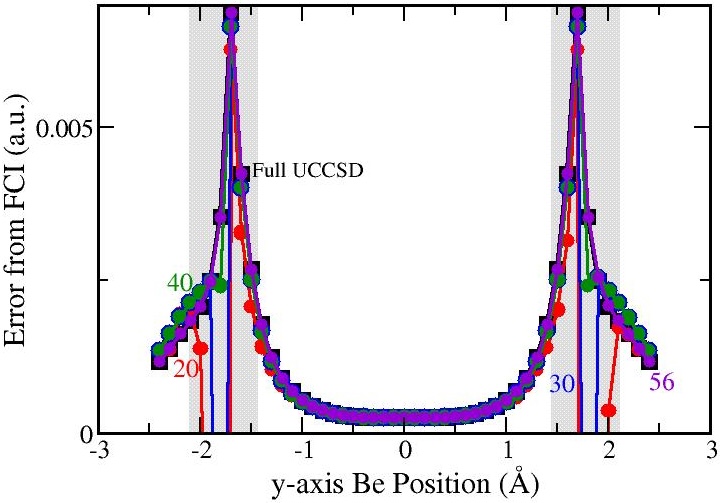} }}

    \caption{{\em Left}: Total energy for the BeH$_2$ system as the beryllium atom is inserted. {\em Right}: The error from FCI, as $E-E_\text{FCI}$, for several different methods. Here, 40 exact factors are used for qUCCSD and we see that already, it performs similar to both UCCSD and CCSD, and outperforms CCSD in the region of strong correlations. {\em Bottom}: Energy difference as we add more factors to achieve convergence. We find that with 56 factors (40 doubles and 16 singles), qUCCSD recovers almost the full UCCSD with the order of one-fourth of the UCC factors. In all panels, the gray region denotes where the single reference state produces the first excited state.  }
    \label{fig:BeH2_1}
\end{figure}

The results are summarized in Figure~\ref{fig:BeH2_1}. As before, we determine which angles initially to treat as large from the MP2 values. We see that the qUCCSD immediately provides accurate results for the weakly correlated regimes. We also study qUCCSD for different numbers of large angles to seek convergence for the strong correlation regime. We see that there also seems to be a troublesome intermediate regime, but we find that by including just 40 doubles and 16 singles for a total of 56 large factors, we recover almost exactly the result of the full UCCSD, which has 206 factors. Moreover, qUCCSD outperforms CCSD in the  regime where the conventional CC gives poor results. Thus, with a careful selection of singles and doubles together, we achieve accuracy on par with the full UCCSD using about a quarter of the total factors. This again illustrates that the qUCCSD idea is working well even for a markedly different type of system.   

\section{Conclusion}

Electronic structure of molecules has long been described as the ``killer application'' for quantum computers~\cite{quantcomp}. But, closer inspection shows that current hardware requires very low depth circuits to be able to work well~\cite{NISQ,qUCC_imp,trapped_ion_sim, dejong, ionq, UCC_imp,UCC_imp2}. By introducing an algorithm where only a small fraction of UCC factors are applied to the HF reference on the quantum computer and the remaining terms are treated using a Taylor-series expansion about the origin for the small angles and the exact values for the large angles, allows the circuits to be substantially smaller. We find that by systematically increasing the number of UCC factors included exactly, we can achieve convergence to the full UCCSD result. For the same accuracy as UCCSD, we typically need between 25\% and 35\% of the UCC factors to actually be put onto the quantum computer. This is a significant savings and a significant reduction in the number of factors needed to be put onto the quantum computer. 
The qUCC method provides this circuit reduction by limiting the number of factors needed to be treated exactly and the number of exact factors needed appears to grow slower than the system size (although we have only examined a small number of systems so cannot make any strong conclusions about this).  There is good reason to anticipate that this trend will continue for even larger systems, as the number of factors with large amplitudes is known to be a small subset of all of the amplitudes from other methods such as MP2 and CC. Our work also shows that the convergence is most challenging in the crossover region between weak and strong coupling, with rapid convergence (and a small number of exact factors needed) for both extremes. This is a quite favorable result for generating accurate data for wide regions of parameter space.

\section{Acknowledgments}
J. K. F. and J. C. were supported by the National Science Foundation under grant number CHEM-2154671. D. Z. was supported by the National Science Foundation under grant number CHEM-2154672.
 J.K.F. was also supported by the
McDevitt bequest at Georgetown University. We acknowledge technical assistance from Zach He (Georgetown) Vibin Abraham (Michigan) in running the quantum chemistry codes.

\end{document}